# Fundamental Efficiency Limits for Small Metallic Antennas

Carl Pfeiffer, *Member, IEEE*

*Abstract* – Both the radiation efficiency and bandwidth of electrically small antennas are dramatically reduced as the size decreases. Fundamental limitations on the bandwidth of small antennas have been thoroughly treated in the past. However, upper bounds on radiation efficiency have not been established even though it is also of significant importance. Here, radiation from a thin metallic shell is rigorously analyzed to establish fundamental limits on the radiation efficiency of resonant, electrically small antennas in terms of the size and the metal conductivity. Metallic losses are systematically introduced into the circuit model proposed by Chu, and several resonant antennas with maximum radiation efficiencies are analyzed. Resonant electric and magnetic dipole antennas both have maximum radiation efficiencies near 100% until the size is reduced below a critical value, at which point the efficiency scales as electrical size to the fourth power ($(ka)^4$). It is also shown that a helix antenna that resonantly couples the $TM_{10}$ mode to the $TE_{10}$ mode has a maximum radiation efficiency, and is about twice that of a resonant dipole or loop antenna. The closed form expressions reported here provide valuable insight into the design of small antennas with optimal efficiencies.

## I. INTRODUCTION

THERE is a continual desire to reduce the size of antennas for wireless communication. However, it is well-known that the radiation efficiency ($e_{rad}$) and bandwidth ($B$) of electrically small antennas are dramatically reduced as the size decreases. In the 1940's, Wheeler and Chu established a lower bound on the minimum radiation quality factor ($Q_{rad}$) that a small antenna can realize [1, 2], which is related to the antenna's bandwidth [3]. This pioneering work prompted tremendous research efforts towards developing small antennas with bandwidths approaching these fundamental limits [4]. Furthermore, a myriad of more accurate lower bounds on the quality factors of antennas with different geometries, material loadings, radiation patterns, etc. were also established [5].

However, the communication data rate, rather than bandwidth, is often the most meaningful metric for characterizing the overall performance of a wireless link. The maximum rate that data can be transmitted is known as the channel capacity ($C$), and is related to the bandwidth and signal-to-noise ratio ($S/N$) at the receiver by [6],

$$C = B \log_2 \left(1 + \frac{S}{N}\right), \quad (1)$$

Note that the signal-to-noise ratio of omnidirectional antennas is proportional to the radiation efficiency. Communication systems commonly realize data rates that closely approach the channel capacity [7].

Let us consider some of the implications of (1) on the design of electrically small antennas with omnidirectional radiation patterns. It is well known that the bandwidth of a small antenna can be increased by resistive loading, which reduces the radiation efficiency. This tradeoff is particularly useful when there is a large signal-to-noise ratio ($S/N \gg 1$) at the receiver such that the net effect of increasing $B$ and reducing $S/N$ enhances the channel capacity, $C$. However, there are also situations where the signal-to-noise ratio is low (e.g., GPS, wireless sensor networks, internet of things, implantable antennas) [8-10]. When $S/N \ll 1$, the channel capacity can be written as,

$$\lim_{S/N \to 0} C = \frac{S}{\ln(2) N_0}, \quad (2)$$

where $N_0 = N/B$ is the power spectral density from a white noise source. In this extreme scenario, the channel capacity is independent of the bandwidth since increasing bandwidth also increases noise. From an antenna perspective, the maximum data rate of these power limited systems can only be improved by increasing the radiation efficiency. Clearly, it is important to quantify tradeoffs between size, efficiency, and bandwidth of electrically small antennas so that wireless systems with optimal performances can be designed. However, the vast majority of the small antenna literature has only focused on tradeoffs between bandwidth and size.

Small antennas can typically be categorized as either an electric dipole ($TM_{10}$) or a magnetic dipole ($TE_{10}$). Electric dipoles have substantially larger radiation resistances than magnetic dipoles, but they are also more difficult to match. It is an open question as to which antenna has a better radiation efficiency when tuned to resonance. To date, the most common method of analytically estimating tradeoffs between radiation efficiency and size is to consider common antenna types (e.g. capacitively loaded loop, inductively loaded dipole), and then approximate the current distribution to calculate the radiated

The author is an on-site contractor with Defense Engineering Corporation at Air Force Research Laboratory, Sensors Directorate, 2241 Avionics Circle, Wright-Patterson AFB, OH 45433-7320 (email: carlpfei@umich.edu).

and dissipated power [11-13]. However, this approach is not general.

There have been some attempts to establish fundamental efficiency limits. In [14], the efficiency of a small antenna is written in terms of the efficiency of the matching network and the radiation efficiency of the antenna. However, limited insight can be gained by treating the antenna and matching network independently since it is usually more efficient to integrate the matching network into the antenna itself [15]. An attempt to establish bounds on the maximum achievable gain and efficiency using a "loss merit factor" is reported in [16]. However, the results are clearly unphysical since a single turn loop antenna can surpass these fundamental limits when $ka \ll 1$ [12]. The most general characterization of the radiation efficiency was performed by Harrington, who established fundamental limits on the gain, bandwidth, and efficiency of arbitrarily sized antennas [17]. However, the efficiency and bandwidth are written in general terms of spherical Bessel functions, and are not simplified for the case of electrically small antennas. Furthermore, there is no discussion on how to establish a resonance, which is a necessary condition for impedance matching. In [17], Harrington derived the familiar expression for the maximum achievable gain of a given sized antenna $(G = (ka)^2 + 2ka)$, and notes that it is only valid when the antenna is electrically large $(ka > 1)$. However, the formula is commonly misapplied to the case of electrically small antennas [18-21].

Here, the circuit model introduced by Chu is used to derive fundamental limitations on the radiation efficiency of small antennas. To begin, the analysis proposed by Harrington is examined in detail for the case of electrically small antennas. Then, more accurate bounds are established that account for energy stored within the antenna, while also enforcing a resonance. It is postulated that a thin metallic shell represents a physical structure with maximal radiation efficiency. Then, radiation from the shell is rigorously analyzed, which is possible due to its spherical symmetry. It is shown that the maximum radiation efficiency of a small antenna is a function of the electrical size $(ka)$, operating frequency, and metal conductivity. This work provides a rigorous standard that can be used to characterize the relative performance of different small antenna designs. Furthermore, the insight provided here can aid in designing antennas with optimal radiation efficiency for a given size.

## II. Definitions and Assumptions

To begin, it is useful to outline some definitions and the assumptions that are made. The quality factor is the ratio of the energy stored $(W)$ to the power dissipated $(P)$. Since power can be dissipated through either radiation $(P_{rad})$ or material losses $(P_{loss})$, it is useful to define two different quality factors. The radiation quality factor is defined as,

$$Q_{rad} = \frac{2\omega W}{P_{rad}}, \qquad (3)$$

which neglects material losses. Lower bounds on the radiation quality factor have been thoroughly treated in the past. The total quality factor,

$$Q_{tot} = \frac{2\omega W}{P_{tot}} = \frac{2\omega W}{P_{rad} + P_{loss}} = e_{rad} Q_{rad}, \qquad (4)$$

accounts for both radiation and material losses, where $e_{rad}$ is the radiation efficiency. In both cases, $W$ represents the larger of the electric $(W_e)$ or magnetic $(W_m)$ stored energy, and $\omega$ is the angular frequency.

It is often more convenient to discuss the dissipation factor, $D$, rather than the radiation efficiency [17]. The dissipation factor is defined as the ratio of the power lost to material absorption to the power radiated,

$$D = \frac{P_{loss}}{P_{rad}}, \qquad (5)$$

and is related to the radiation efficiency by,

$$e_{rad} = \frac{P_{rad}}{P_{rad} + P_{loss}} = \frac{1}{1 + D}. \qquad (6)$$

Since there exists a one-to-one relationship between the dissipation factor and the radiation efficiency, the two terms are used interchangeably.

The limitations on radiation efficiency derived here are valid for metallic antennas with metal thicknesses that are much larger than the skin depth. In general, the efficiency of the antenna increases if it is filled with a magnetic material with high permeability and sufficiently low loss. If such a material exists, the geometries discussed here are not necessarily optimal. In addition, it is possible for dielectric loaded antennas to achieve resonance using materials with a high permittivity $(> 50)$ and low loss. These antennas may have efficiencies that surpass the fundamental limits derived here. However, since the majority of small antennas operating above 1 MHz do not utilize either magnetic or dielectric loading, it is likely that any potential benefit to the radiation efficiency is offset by increased material losses. A more thorough analysis on the limitations of magneto-dielectric loaded antennas is left for future work.

## III. Radiation from a Solid Metallic Sphere

In [17], Harrington considered radiation from a solid metallic sphere to analytically compute antenna gain limitations. In this section, we review the analysis performed by Harrington using the circuit models introduced by Chu [1] and Thal [22]. In addition, the efficiency bounds are simplified by taking the limit that the size is electrically small.

The circuit model shown in Fig. 1 was proposed in [22] to exactly model the modes supported by a spherical geometry with radius $a$, where $c$ is the velocity of light in free space. The circuit model can be derived using the recurrence relations of spherical Bessel functions. All impedances are normalized to the wave impedance of free-space, $\eta_0 = \sqrt{\mu_0/\varepsilon_0}$. At each terminal, the impedance looking to the left represents the wave impedance of outward propagating spherical modes, whereas the impedance looking to the right represents the wave impedance of inward propagating modes. The total energy stored in the electric and magnetic fields can be calculated by summing the energy stored in all capacitors ($W_e = \sum CV^2/4$) and inductors ($W_m = \sum LI^2/4$), respectively. The circuit is valid for all indices $m$. The TM$_{1m}$ modes have radiation patterns identical to short electric dipoles, whereas the TE$_{1m}$ modes have radiation patterns identical to small loops (i.e. magnetic dipoles). For simplicity, it will be assumed that $m = 0$. It is important to note that all TM and TE spherical modes are orthogonal, which allows each mode to be treated independently of the others.

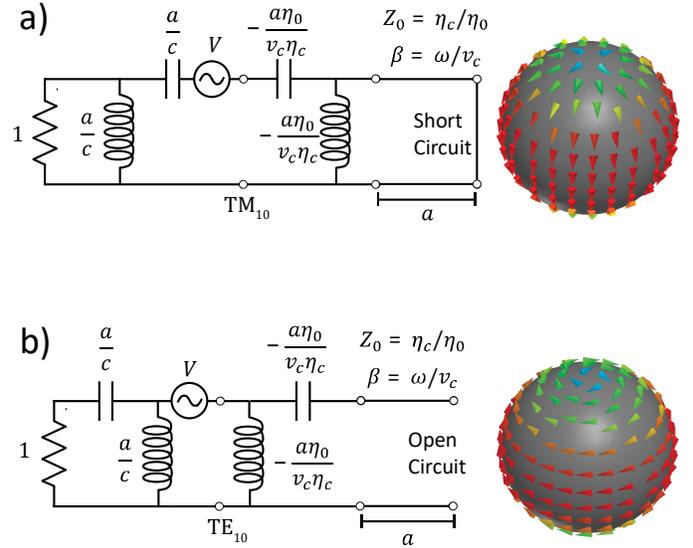

Fig. 2: Alternate form of the circuit shown in Fig. 1 for the case of a metallic sphere that is excited with magnetic current sources directly above its surface. (a) The TM$_{10}$ mode is excited. (b) The TE$_{10}$ mode is excited. The electric surface currents induced on the surface of the metal are shown to the right for the two cases.

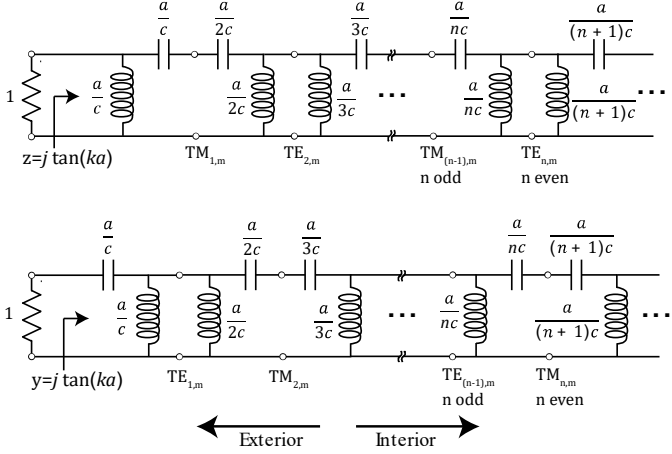

Fig. 1: Exact circuit model introduced by Thal that models spherical modes external and internal to a sphere of radius $a$. All impedances are normalized to the wave impedance of free space, $\eta_0$.

Consider a sphere with radius $a$ that is filled with a highly conductive metal with normalized impedance much less than its electrical size ($\eta_c/\eta_0 \ll ka$), where $k = \omega/c$ is the wavenumber of free space. The normalized impedance of the conductor can be related to the conductivity ($\sigma$) and skin depth ($\delta_s$) by,

$$\frac{\eta_c}{\eta_0} \approx (1+j)\sqrt{\frac{\omega\varepsilon_0}{2\sigma}} = \frac{k\delta_s}{2}. \quad (7)$$

If the conductivity is large, the reactive component of $\eta_c/\eta_0$ can be neglected since it will be dominated by the other reactive elements in the circuit. This hypothetical antenna is excited with magnetic current sources directly above the surface that excite the lowest order modes (TM$_{10}$ and TE$_{10}$) since these radiate the most efficiently. These represent the most ideal conditions for maximizing the efficiency since any structuring of the metal will only increase the dissipation factor ($D = P_{loss}/P_{rad}$).

The sphere excited with impressed magnetic current sources can be modelled using the circuits shown in Fig. 2 (a) and (b) for the TM$_{10}$ and TE$_{10}$ modes, respectively. Note that the circuit model is an alternative form of the circuit shown in Fig. 1, and is terminated with a transmission line of length $a$. It is sometimes more convenient to use this circuit model, which can be derived by noting that the impedance or admittance looking inwards from the free space resistance is equal to $j\tan(ka)$ for the two cases, respectively [22].

If the metal has a high conductivity, the negative inductor and negative capacitor to the right of the voltage source become open and short circuits, respectively. Therefore, the circuit can be simplified as shown in Fig. 3 for the case of electrically small antennas ($ka < 0.5$). Fundamental limits on the radiation efficiency of small electric and magnetic dipole antennas can be calculated from these circuit models. The dissipation factors for the TM$_{10}$ and TE$_{10}$ modes are,

$$D_{ideal}^{TM10} = \frac{(\eta_c/\eta_0)}{(ka)^2} + O((ka)^0),$$
$$D_{ideal}^{TE10} = \frac{(\eta_c/\eta_0)}{(ka)^4} + O\left(\frac{1}{(ka)^2}\right), \quad (8)$$

where $O((ka)^n)$ designates terms that have an order of magnitude $(ka)^n$. These dissipation factors qualitatively agree with previous analyses that considered the radiation efficiency of short dipoles and small loops [11]. Electric dipoles have a much larger input resistance and therefore a larger efficiency for a given sized antenna. However, this analysis neglects impedance matching since a solid conducting sphere cannot resonate, which is a necessary condition for impedance matching high $Q$ antennas. Furthermore, this analysis also neglects energy stored within the antenna.

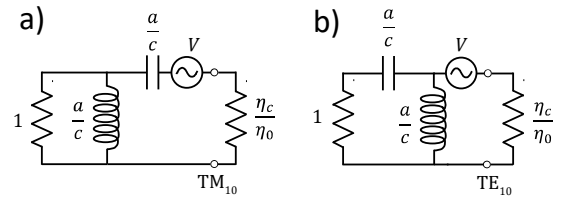

Fig. 3: Simplified circuits that model radiation from a highly conductive, electrically small sphere. (a) TM$_{10}$ radiation. (b) TE$_{10}$ radiation.

## IV. RADIATION FROM A METALLIC SHELL

It is postulated that radiation from a thin metallic shell provides an upper limit on the radiation efficiency of resonant antennas. Consider a metallic shell with radius $a$ and thickness $t$, as shown in Fig. 4. All realistic antennas pattern the metal to establish a resonance, which only increases metallic losses. The metallic shell is filled with a material with relative permittivity $\varepsilon_s$ and relative permeability $\mu_s$, which are assumed to be lossless for simplicity. Placing any metal within the sphere will only reduce the radiation efficiency since it will increase the energy stored within the antenna (and therefore dissipation) without adding radiated power.

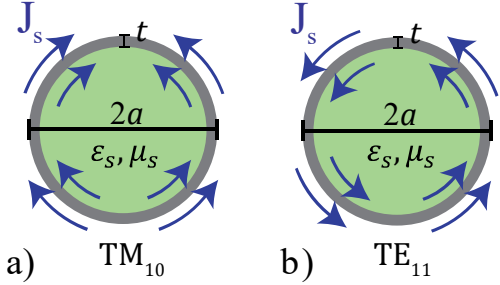

Fig. 4: A metallic shell with thickness $t$ is filled with a material with relative permittivity $\varepsilon_s$ and relative permeability $\mu_s$. (a) and (b) Induced surface currents generating TM$_{10}$ radiation and TE$_{11}$ radiation, respectively.

The circuits that exactly model TM$_{10}$ and TE$_{10}$ radiation are shown in Fig. 5 [22]. Again, the metallic shell is assumed to have a high conductivity ($\eta_c/\eta_0 \ll ka$), which allows it to be replaced with a transmission line of length $t$ and normalized impedance $\eta_c/\eta_0$.

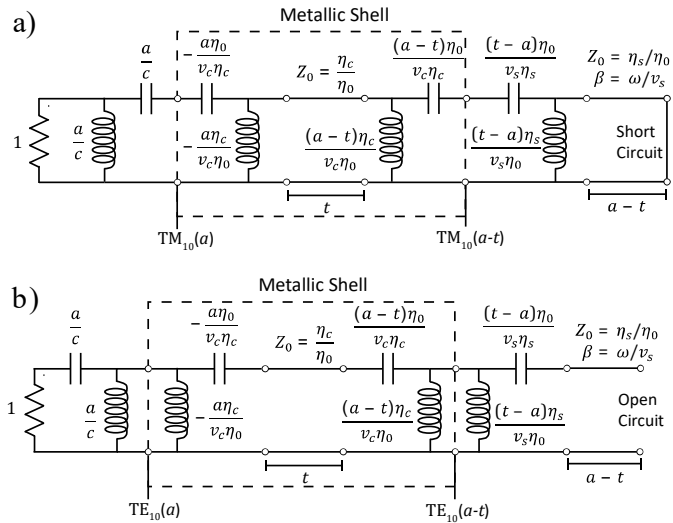

Fig. 5: (a) and (b) Exact circuits corresponding to a metallic shell that radiates the TM$_{10}$ and TE$_{10}$ modes, respectively.

The shell thickness is assumed to be much greater than the skin depth of the metal, but much less than the radius ($\delta_s \ll t \ll a$). Therefore, the shell can be replaced with two resistors $\eta_c/\eta_0$ that isolate the external field from the internal field, as shown in Fig. 6. In addition, the antenna is assumed to be electrically small such that the impedance looking into the origin is represented as a series or parallel LC circuit for the TM or TE cases, respectively. The antenna is symmetrically fed with magnetic current sources on either side of the metallic shell, which excite the lowest order TM or TE modes for optimal radiation efficiency. This excitation ensures the voltage on the exterior surface of the antenna ($V^+$) is identical to the voltage on the interior surface ($V^-$), at every position ($V^+(\theta,\phi) = V^-(\theta,\phi)$). In other words, the exterior surface of the metallic shell is effectively shorted to the interior surface of the shell, which is the case for all realistic antennas utilizing metal patterned on a spherical surface.

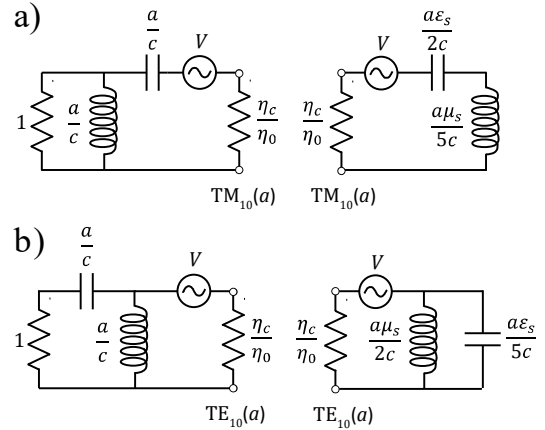

Fig. 6: (a) and (b) Approximate circuits modelling TM$_{10}$ and TE$_{10}$ radiation from a metallic shell for the case where $\eta_c/\eta_0 \ll kt \ll ka \ll 1$, respectively. For both cases, the voltage sources symmetrically excite the inside and outside of the metallic shell such that the exterior surface of the shell is effectively shorted to the interior surface.

From symmetry, the circuits shown in Fig. 6 can be further simplified to the circuits shown in Fig. 7. The dissipation factors for the TM$_{10}$ and TE$_{10}$ modes can be calculated using the derived circuit,

$$D_{shell}^{TM10} = \frac{(1+(\varepsilon_s/2)^2)(\eta_c/\eta_0)}{(ka)^2} + O((ka)^0),$$
$$D_{shell}^{TE10} = \frac{(1+(2/\mu_s)^2)(\eta_c/\eta_0)}{(ka)^4} + O((ka)^0). \quad (9)$$

If the antenna is loaded with a perfect magnetic conductor ($\mu_s \to \infty, \varepsilon_s \to 0$), the dissipation factors reduce to the case analyzed by Harrington. In this case, no energy is stored within the antenna since the internal inductor and capacitor are open circuited. Eq. (9) clearly shows that the radiation efficiency is reduced as the permeability is decreased and the permittivity is increased.

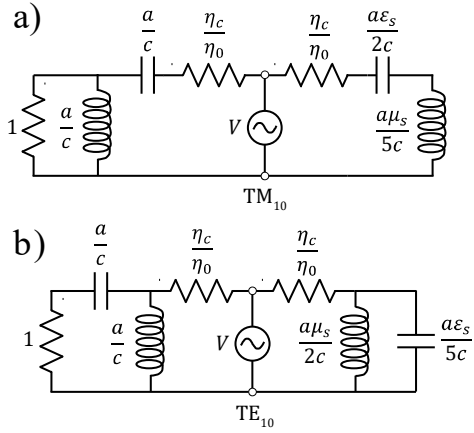

Fig. 7: Simplified version of the circuit shown in Fig. 6.

It should be noted that the circuits shown in Fig. 7 could have been directly written using a different argument that does not require considering the intermediate steps of Fig. 5 and Fig. 6. A small spherical antenna can be replaced with impressed electric current sources over a spherical surface. This scenario is modelled by replacing the shunt voltage source in Fig. 7 with a shunt current source, as shown in [23]. Then, the metallic loss is found by integrating the quantity $|H_t|^2 \eta_c$ over the surface of the antenna, where $H_t$ represents the tangential magnetic field (i.e., the surface current). In this circuit model, the magnetic field next to the surface of impressed current sources is represented as current flowing through the two resistors with impedance $\eta_c/\eta_0$. Therefore, these two resistors are required to account for dissipation on the two metal surfaces, which exist on the external and internal sides of the sphere.

There are several reasons why the approach using impressed magnetic current sources is emphasized here. This work is closely related to the analysis used by Harrington [17]. Therefore, it is natural to use the same method that Harrington employed. In addition, the approach seemed to be more physical since a geometry (spherical metallic shell) is first considered, and then sources are added to generate radiation. This systematic methodology makes it clear which assumptions are made ($\eta_c/\eta_0 \ll ka$ and $\delta_s \ll t \ll a$). The reason why two $\eta_c/\eta_0$ resistors are required to model metallic losses rather than one resistor in series with the generator is subtle. The magnitude of the surface currents flowing on the internal and external surfaces of the metallic shell are in general different, and therefore requires two separate resistors. This outcome is a natural result when impressed magnetic currents over a spherical shell are considered, and is less obvious if impressed electric current sources are used.

*A. Quality Factor Limitations for Capacitors*

These circuit models can also be used to derive upper bounds on the performance of inductors and capacitors. In the limit $ka \to 0$, the circuit shown in Fig. 7(a) simplifies to a lossy capacitor (series RC circuit), with quality factor,

$$Q_{tot}^C = \frac{2(2+\varepsilon_s)}{4(ka)^3 + (\varepsilon_s^2 + 4)(ka)(\eta_c/\eta_0)} + O\left(\frac{1}{ka}\right). \quad (10)$$

The term proportional to $(ka)^3$ corresponds to radiative loss and the term proportional to $(\eta_c/\eta_0)$ corresponds to ohmic loss. As the size of the capacitor decreases, the $Q$ increases. Furthermore, the Q increases as the excited mode order $(n)$ increases. Therefore, there is no fundamental limit on the quality factor of a capacitor constructed from metal. In practice, capacitors can have extremely large Q factors (exceeding 1000), which are generally limited due to dielectric breakdown or manufacturing capabilities of accurately fabricating small gaps.

*B. Quality Factor Limitations for Inductors*

Conversely, metallic losses provide an upper bound on the achievable quality factor of an inductor. In the limit $ka \to 0$, the circuit shown in Fig. 7(b) simplifies to a lossy inductor (series RL circuit). The $Q$ of this circuit is,

$$Q_{tot}^L = \frac{\mu_s(2+\mu_s)(ka)}{\mu_s^2(ka)^4 + (4+\mu_s^2)(\eta_c/\eta_0)} + O\left(\frac{1}{ka}\right). \quad (11)$$

The term proportional to $(ka)^4$ corresponds to radiative loss, whereas the term proportional to $(\eta_c/\eta_0)$ corresponds to metallic loss. If higher order modes are excited, the metallic loss remains the same, but the inductance is decreased. Therefore, exciting the TE$_{10}$ mode provides an optimal Q for inductors provided that $(ka)^4 \ll 1$ such that the radiative loss is not significant. In other words, an arbitrary inductor within a sphere of radius $a$ will always have a quality factor lower than (11). It should be noted that the inductor Q in (11) is expressed in terms of the magnetic permeability $\mu_s$ for completeness. However, if a material with high permeability and sufficiently low loss is available, the metallic shell shown in Fig. 4 is not the optimal configuration for maximizing the inductor Q. One example that is more efficient is a metallic shell with radius $a/2$ embedded within a magnetic material with radius $a$. Nevertheless, it seems that existing materials typically have loss tangents that are prohibitively large above 1 MHz since commercially available inductors with the highest quality factors use air cores [24].

## V. RESONANT SMALL ANTENNAS

In order to impedance match a small antenna to a resistive load, it must resonate to cancel the input reactance. The three most efficient resonant antennas will be considered: capacitively loaded TE$_{10}$ antenna, coupled TM$_{10}$:TE$_{10}$ antenna, and coupled TM$_{10}$:TE$_{20}$ antenna.

*A. Capacitively loaded TE$_{10}$ Antenna (Magnetic Dipole)*

A loop antenna radiates the TE$_{10}$ mode. The antenna will resonate if a capacitor is placed in series with the feed. As demonstrated in the previous section, it is reasonable to assume that the capacitor has an infinitely large $Q_{tot}$ when deriving an upper bound on the radiation efficiency. Adding an ideal capacitor in series with the voltage source in Fig. 7(b) does not affect the radiation efficiency. Therefore, $D_{res}^{TE10}$ of a resonant loop antenna is equal to $D_{shell}^{TE10}$. If $\varepsilon_s = 1$ and $\mu_s = 1$, the dissipation factor simplifies,

$$D_{res}^{TE10} = \frac{5}{(ka)^4}\left(\frac{\eta_c}{\eta_0}\right) + \frac{11}{5(ka)^2}\left(\frac{\eta_c}{\eta_0}\right) + O((ka)^0). \quad (12)$$

It should be noted that the radiation efficiency of a loop antenna is often written in terms of the number of turns $N$, which does not appear in (12). A multi-turn loop antenna has an inductance and radiation resistance that increases as $N^2$. However, the loss resistance only increases as $N$, which suggests that increasing the number of turns increases the radiation efficiency [25]. However, this argument assumes that the wire diameter is independent of $N$. In principle, a single-turn antenna could have a wire diameter that is $N$ times larger than a multi-turn antenna with the same overall dimensions, which reduces the loss resistance. Therefore, single-turn antennas and multi-turn antennas theoretically have similar radiation efficiencies, which agrees with (12). In practice, multi-turn loop antennas provide a significantly larger input impedance which makes them easier to match to 50 Ω loads. Of course, these arguments are overly simplistic since they assume the current is uniformly distributed around the wire. In reality, eddy currents cause the current to bunch around the wire's surface, which increases the loss resistance of both multi-turn and single-turn loops [26]. This makes it difficult to accurately estimate the radiation efficiency of wire antennas using conventional methods [11-13].

*B. Coupled $TM_{10}$:$TE_{10}$ Antenna (Self-Resonant Helix)*

Electric dipole antennas ($TM_{10}$) require an inductor to generate a resonance. However, since ideal inductors do not exist, it is not physically meaningful to discuss the dissipation factor of a resonant electric dipole on its own. Instead, the $TM_{10}$ mode should be coupled to either the $TE_{10}$ or $TE_{20}$ modes to establish a resonance with maximal radiation efficiency [23]. First, let us consider coupling between the $TM_{10}$ mode and the $TE_{10}$ mode using an ideal transformer, as shown in Fig. 8. Physically, the circuit corresponds to directly exciting the $TE_{10}$ and $TM_{10}$ modes with magnetic currents directly above the surfaces of the metallic shell with a ratio of $1:N$ to ensure resonance ($W_e = W_m$). In practice, it is also possible to directly excite both the $TE_{10}$ and $TM_{10}$ modes using a properly designed antenna. For instance, the self-resonant spherical helix antenna proposed by Wheeler efficiently couples the two modes using a helical wire with varying pitch [13]. If $\varepsilon_s = 1$ and $\mu_s = 1$, the number of turns $N$ that are required to establish a resonance simplifies,

$$N = \frac{\sqrt{2}}{(ka)} - \frac{11\sqrt{2}(ka)}{20} + O((ka)^3), \quad (13)$$

as demonstrated by Thal [23]. At resonance the dissipation factor of this coupled system is

$$D_{res}^{TM10:TE10} = \frac{5}{3(ka)^4}\left(\frac{\eta_c}{\eta_0}\right) + \frac{17}{30(ka)^2}\left(\frac{\eta_c}{\eta_0}\right) + O((ka)^0). \quad (14)$$

The dissipation factor is 3 times lower than the dissipation factor of the magnetic dipole on its own. Eq. (14) represents a lower bound on the dissipation factor of an electrically small antenna without material loading. The ratio of the power radiated in the $TM_{10}$ mode to the power radiated in the $TE_{10}$ mode is,

$$\frac{P_{rad}^{TM10}}{P_{rad}^{TE10}} = 2 + \frac{9(ka)^2}{5} + 4\left(\frac{\eta_c}{\eta_0}\right)(ka)^2 + O((ka)^4) \quad (15)$$

Therefore, the radiation pattern is in general elliptically polarized [27].

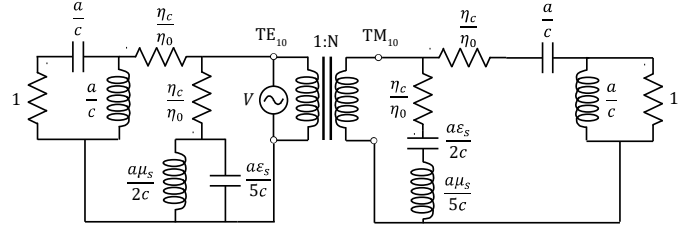

Fig. 8: The $TM_{10}$ mode is coupled to the $TE_{10}$ mode using an ideal transformer to establish a resonance. This circuit models the upper bound on the radiation efficiency of electrically small antennas without material loading.

*C. Coupled $TM_{10}$:$TE_{20}$ Antenna (Electric Dipole)*

If it is desired to have a radiation pattern nearly identical to that of a short electric dipole, the $TM_{10}$ mode can be coupled to the $TE_{20}$ mode, which radiates poorly [23] (see Fig. 9). Physically, this is similar to the case of a spherical helix antenna mounted on top of a ground plane [28]. However, not only does the $TE_{20}$ mode have a lower radiation resistance $\left(2P_{rad}/\left|I_{feed}\right|^2\right)$, it also has a lower inductance (i.e. lower $Q_{tot}^L$), which increases the dissipation factor further. If $\varepsilon_s = 1$ and $\mu_s = 1$, the number of turns $N$ that are required to establish a resonance simplifies,

$$N = \frac{\sqrt{10/3}}{(ka)} - \frac{167\sqrt{30}(ka)}{1260} + O((ka)^3). \quad (16)$$

At resonance, the dissipation factor of this coupled system is

$$D_{res}^{TM10:TE20} = \frac{39}{10(ka)^4}\left(\frac{\eta_c}{\eta_0}\right) - \frac{47}{140(ka)^2}\left(\frac{\eta_c}{\eta_0}\right) + O((ka)^0), \quad (17)$$

and the ratio of the power radiated in the $TM_{10}$ mode to the power radiated in the $TE_{20}$ mode is

$$\frac{P_{rad}^{TM10}}{P_{rad}^{TE20}} = \frac{30}{(ka)^2} + \frac{113}{7} + O((ka)^2). \quad (18)$$

Therefore, the radiation pattern is predominantly that of an electric dipole when the antenna is electrically small.

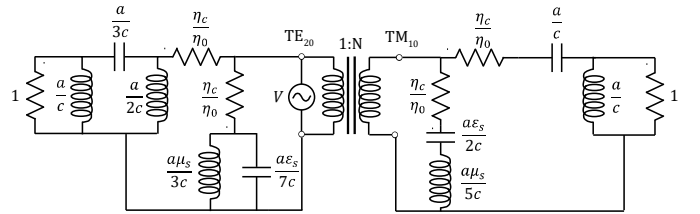

Fig. 9: The $TM_{10}$ mode is coupled to the $TE_{20}$ mode using an ideal transformer to establish a resonance. This system has a radiation pattern virtually identical to that of a short electric dipole since the $TE_{20}$ does not radiate efficiently.

The dissipation factor of this dipole antenna is ~2.3 times larger than the case where there is $TM_{10}$:$TE_{10}$ coupling (i.e. the self-resonant helix antenna). Even though electric dipole antennas require lossy inductors to be impedance matched, they have a substantially larger radiation resistance than loop antennas. These two competing factors end up cancelling each other, such that that the overall radiation efficiency of dipole and loop antennas are quite similar ($D_{res}^{TM10:TE20}/D_{res}^{TE10} \approx 0.78$). This agrees with conventional wisdom since both electric and magnetic dipoles are commonly utilized.

Table 1 summarizes the main results of this section. The leading order terms of the quality and dissipation factors for each antenna are reported for the case where $\varepsilon_s = 1$ and $\mu_s = 1$. All expressions are simplified by noting that $k\delta_s/2 = \eta_c/\eta_0$. In addition, metallic geometries with similar radiation properties are shown in the left column. It should be emphasized that these metallic geometries do not correspond to optimized antennas, and are only provided to give some physical intuition. The TM$_{10}$ antenna supports $\hat{\theta}$-directed currents along its surface, but must be matched with an ideal inductor, which is unphysical. The TE$_{10}$ and TE$_{20}$ antennas are each fed with a capacitively loaded vertical wire through the center, so that current can continually flow around a loop. The TM$_{10}$:TE$_{10}$ and TM$_{10}$:TE$_{20}$ coupled antennas are self-resonant, and do not require lumped element loading.

Table 1: Quality and dissipation factors for the different antenna types when $\varepsilon_s = 1$ and $\mu_s = 1$. The dissipation factor of the TM$_{10}$:TE$_{10}$ coupled antenna establishes a fundamental limit for small antennas without material loading. All expressions are simplified by noting that $k\delta_s/2 = \eta_c/\eta_0$.

| | $Q_{tot}$ | $D_{res}$ | Notes |
|---|---|---|---|
| TM$_{10}$ 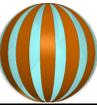 | $\dfrac{3/2 + 12(ka)^2}{(ka)^3 + \left(\dfrac{5}{8}\right)(k^2\delta_s a)}$ | $\dfrac{5}{8ka}\left(\dfrac{\delta_s}{a}\right)$ | Tuned with an ideal inductor (nonphysical) |
| TE$_{10}$ 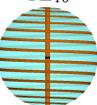 | $\dfrac{3 + 3(ka)^2}{(ka)^3 + \dfrac{5\delta_s}{2a}}$ | $\dfrac{5}{2(ka)^3}\left(\dfrac{\delta_s}{a}\right)$ | Tuned with an ideal capacitor |
| TE$_{20}$ 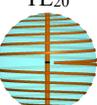 | $\dfrac{45 + 15(ka)^2}{(ka)^5 + \dfrac{117\delta_s}{2a}}$ | $\dfrac{117}{2(ka)^5}\left(\dfrac{\delta_s}{a}\right)$ | Tuned with an ideal capacitor |
| TM$_{10}$:TE$_{10}$ 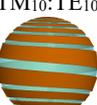 | $\dfrac{1 + 11(ka)^2/10}{(ka)^3 + \dfrac{5\delta_s}{6a}}$ | $\dfrac{5}{6(ka)^3}\left(\dfrac{\delta_s}{a}\right)$ | Self-resonant (maximum efficiency) |
| TM$_{10}$:TE$_{20}$ 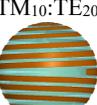 | $\dfrac{3/2 + 97(ka)^2/140}{(ka)^3 + \dfrac{39\delta_s}{20a}}$ | $\dfrac{39}{20(ka)^3}\left(\dfrac{\delta_s}{a}\right)$ | Self-resonant |

## VI. DISCUSSION

The radiation efficiency of a small antenna depends upon the surface resistivity of the metal ($\eta_c/\eta_0$), which is a function of the operating frequency and conductivity. Upper bounds on the radiation efficiency of the three most efficient antenna types constructed from copper ($\sigma = 5.96 \times 10^7$ S/m) that operate at 300 MHz are shown in Fig. 10. The radiation efficiency rapidly deteriorates when the antenna is smaller than the critical size,

$$ka_{critical} \sim (k\delta_s)^{1/4} = \left(\dfrac{2\omega\varepsilon_0}{\sigma}\right)^{1/8}, \quad (19)$$

which equals 0.07 for copper at 300 MHz. The dissipation factor is near unity at this size. It should be noted that this critical size varies slowly with changes to the operating frequency and conductivity (to the 1/8 power). For example, when the operating frequency is increased by a factor of 100 (30 GHz), the critical size of an antenna constructed from copper is $ka_{critical} = 0.12$.

The dissipation factor of a small antenna is approximately equal to the minimum radiation quality factor multiplied by the normalized skin depth,

$$D \approx Q_{rad}^{min}\left(\dfrac{\delta_s}{a}\right). \quad (20)$$

The minimum quality factors for the three different antennas are $Q_{rad}^{min} = 3/(ka)^3$ for magnetic dipoles, $Q_{rad}^{min} = 1/(ka)^3$ for self-resonant helix antennas, and $Q_{rad}^{min} = 1.5/(ka)^3$ for electric dipoles [23]. The half-power bandwidth of a small antenna can also be written in terms of the radiation quality factor and normalized skin depth,

$$BW_{3dB} \approx \dfrac{2}{Q_{rad}^{min}} + \dfrac{2\delta_s}{a}. \quad (21)$$

When $a > a_{critical}$, the radiation efficiency is near 100%, and the fractional bandwidth is proportional to $a^3$. However, when $a < a_{critical}$ the fractional bandwidth is proportional to $1/a$. In other words, reducing the antenna size actually increases bandwidth when the efficiency is low. This is a due to the fact that the dissipation factor varies more rapidly with electrical size than the radiation quality factor.

The fundamental limits reported here are compared to existing antennas in the literature. The squares and triangles in Fig. 10 correspond to previously published antenna designs, which are spherical and non-spherical, respectively [27, 29-39]. There was an attempt to find antennas with the highest efficiencies, but the list is not comprehensive. The data in Fig. 10 corresponds to both measurement and simulation results. However, simulations are only reported whenever measurements are unavailable. The majority of published antennas have electrical sizes $ka > 0.3$, which are not included. Simulation and measurement errors make it difficult to accurately compare these antennas to the fundamental limit since both are near 100%. To provide a fair comparison between the different antenna designs, the dissipation factors of the published antennas are scaled to account for their different skin depths. Specifically, the dissipation factor is scaled by $(\sqrt{\sigma/f})\sqrt{(300\text{MHz})/(5.96 \times 10^7 \text{ S/m})}$, where $\sigma$ and $f$ are the conductivity and operating frequency of the published antennas. The non-spherical antennas are approximated as spheres with identical volume to provide a fair comparison. In other words, the effective radius of a nonspherical antenna is assumed to be $a = (3V/(4\pi))^{1/3}$, where $V$ is the antenna volume.

There is a great deal of spread in the published data since previous antenna designs were not compared to a well-defined efficiency limit. In contrast, lower bounds on the radiation quality factor were established over a half-century ago, and small antennas regularly approach these limits. In general, the antennas with highest efficiencies maximize the electric currents supported along their outer surface, which is also consistent with minimizing the radiation quality factor. In addition, all antennas need some inductive loading to achieve resonance. The efficiency is maximized when the loading inductor maximizes its size within the antenna, which maximizes the inductor quality factor. For example, antennas that primarily radiate the electric dipole mode should wind a wire around the surface a number of times to generate a large inductance that resonates with the small input capacitance of the $TM_{10}$ mode. Larger wire diameters increase the area that current can flow over, which reduces metallic losses. However, increasing the diameter also reduces spacing between conductors, which in turn increases loss due to the proximity effect [26]. Therefore, an optimal wire diameter should be utilized to minimize the peak current density. Electrically small magnetic dipoles are typically realized with capacitively loaded, single or multi-turn loops. Commercially available, surface mount capacitors commonly realize low loss and high capacitances within small form factors. Therefore, it is relatively straightforward to design magnetic dipole antennas with extremely small electrical sizes. Again, the diameter of the wire should be optimized to spread the current out as much as possible over the surface of the antenna, which in turn maximizes the radiation efficiency.

## VII. SUMMARY

Fundamental limitations on the radiation efficiency of electrically small antennas are derived by considering radiation from a thin metallic shell. It is shown that resonant electric ($D_{res}^{TM10:TE20} \approx 3.9\eta_c/(\eta_0(ka)^4)$) and magnetic ($D_{res}^{TE10} \approx 5\eta_c/(\eta_0(ka)^4)$) dipoles have similar dissipation factors that are over twice that of a self-resonant helix antenna ($D_{res}^{TM10:TE10} \approx 1.67\eta_c/(\eta_0(ka)^4)$). All three cases require a magnetic dipole moment to either radiate or establish resonance. It is the magnetic dipole moment that fundamentally limits the maximum achievable efficiency.

In the future, more accurate bounds on the radiation efficiency can be derived for arbitrarily shaped antennas [40-43]. Furthermore, limitations on the radiation efficiency of magneto-dielectric loaded antennas can also be derived using a similar analysis, and the results should be compared to the limits of metallic antennas. In addition, the results reported here can be generalized to develop more accurate bounds on the maximum gain of electrically large, superdirective antennas [17]. It was mentioned that it is necessary to create a resonance to impedance match an electrically small antenna, but there was no discussion on how to match the antenna to a particular load. In the future, the maximum Q factors of inductors and capacitors reported here, could be integrated into the matching circuit analysis reported in [14] to examine the effects of controlling the input impedance. In addition, the upper bounds on inductor Q could be utilized to develop efficiency limits on wireless power transfer systems [44]. In summary, the simple relationships between antenna size, radiation efficiency, and metal conductivity derived here provide a clear path towards optimizing future antenna designs. The success in developing antennas with near-optimal radiation quality factors inspires confidence that small antennas can be designed that more closely approach these fundamental efficiency limits.

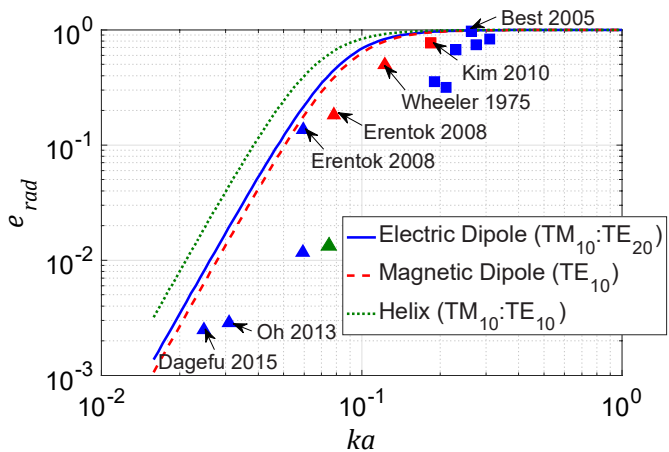

Fig. 10: Radiation efficiency as a function of antenna size. It is assumed the antenna is constructed from copper ($\sigma = 5.96 \times 10^7$ S/m) and operates at 300 MHz ($\eta_c/\eta_0 = 1.18 \times 10^{-5}$). Upper bounds on the electric, magnetic, and helix antennas are shown in blue, red, and green, curves, respectively. Squares and triangles correspond to different antennas in the literature, which are spherical and non-spherical, respectively. Published antennas that have efficiencies approaching the fundamental limit are noted. It is assumed that published antennas that are not spherical have an effective radius ($ka$) equal to that of a sphere with identical overall volume. The dissipation factor of the published antennas is scaled to account for different surface resistivity values. Specifically, the dissipation factor is scaled by $(\sqrt{\sigma/f})\sqrt{(300\text{MHz})/(5.96 \times 10^7 \text{ S/m})}$, where $\sigma$ and $f$ are the conductivity and operating frequency of the published antennas.

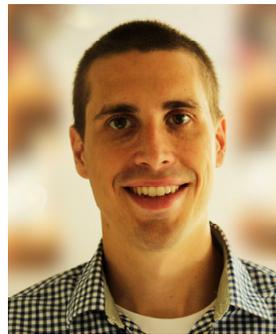

**Carl Pfeiffer** (S'08-M'15) received the B.S.E., M.S.E. and Ph.D. degrees in electrical engineering from The University of Michigan at Ann Arbor, Ann Arbor, MI, USA, in 2009, 2011, and 2015 respectively.

In April 2015, he became a post-doctoral research fellow at The University of Michigan. In March 2016, he joined Defense Engineering Corp. as an onsite contractor for the Air Force Research Laboratory at Wright-Patterson Air Force Base, OH, USA. His research interests include engineered electromagnetic structures (metamaterials, metasurfaces, frequency selective surfaces), antennas, microwave circuits, plasmonics, optics, and analytical electromagnetics/optics.